\date{}
\documentclass[showpacs,superscriptaddress]{revtex4}
\usepackage{graphicx,psfrag,amsmath,amssymb,amsfonts,bbm,latexsym,color,dcolumn,epsf}

\begin{document}
\title{Decoherence induced by zero point fluctuations in quantum Brownian motion}

\author{Fernando C. Lombardo \footnote{lombardo@df.uba.ar}}
\author{Paula I. Villar \footnote{paula@df.uba.ar}}
\affiliation{Departamento de F\'\i sica {\it Juan Jos\'e Giambiagi}, FCEyN UBA,
Facultad de Ciencias Exactas y Naturales, Ciudad Universitaria,
Pabell\' on I, 1428 Buenos Aires, Argentina}

\date{today}

\begin{abstract}

We show a completely analytical approach to the decoherence induced
by a zero temperature environment on a Brownian test particle. We
consider an Omhic environment bilinearly coupled to an oscillator
and compute the master equation. From diffusive coefficients, we
evaluate the decoherence time for the usual quantum Brownian motion
and also for an upside-down oscillator, as a toy model of a quantum
phase transition.

\end{abstract}


\maketitle

\newcommand{\beq}{\begin{equation}}
\newcommand{\eeq}{\end{equation}}
\newcommand{\dalam}{\nabla^2-\partial_t^2}


\section{Introduction}

The emergence of classical behavior from a quantum system is a
problem of interest in many branches of physics \cite{giul}. As it
is well known, the quantum to classical transition involves two
necessary and related conditions: {\it correlations}, i.e. the
Wigner function of a quantum system should have a peak at the
classical trajectories; and {\it decoherence}, that is, there should
be no interference between classical trajectories. To study
quantitatively the emergence of classicality, it is essential to
consider the interaction of the system with its environment, since
both, the loss of quantum coherence and the onset of classical
correlations, depend strongly on this interaction \cite{unruhzu}.
Using this point of view, classicality is an emergent property of an
open quantum system. The strength of the coupling between system and
environment sets the decoherence time which, roughly speaking,
indicates the timescale after which the system can be considered
classical \cite{zurek}.

The very notion of quantum open system implies the appearance of
dissipation and decoherence as an ubiquitous phenomena and plays
important roles in different branches of physics (from quantum field
theory, many body and molecular physics to theory of quantum
information), biology and chemistry. Oftentimes, a large system can
be described adequately as a composite system, consisting of two or
a few subsystems (degrees of freedom) interacting with their
environment (thermal bath) comprising a large number of degrees of
freedom. Examples include electron transfer in solution, large
biological molecules, vibrational relaxation of molecules in
solution, excitons in semiconductors coupled to acoustic or optical
phonon modes. Quantum processes in condensed phases are usually
studied by focusing on a small subset of degrees of freedom and
treating the rest as a bath.

Decoherence is the main ingredient in order to find classicality.
The interaction between the system and the environment induces a
preferred basis which is stable against this interaction, and
becomes a classical basis in the Hilbert space of the coupled
system. Preferred pointer states are resilient to the entangling
interaction with the bath. This ``einselection'' (environment
induced superselection) of the preferred set of resilient pointer
states is the essence of the environment. It is accepted that a
rapid loss of coherence caused by the coupling with the environment
is at the root of the non-observation of quantum superpositions of
macroscopically different quantun states \cite{zurek}. 
A relevant property of the pointer states is their insensitivity to 
being monitored by the interaction with environment 
(and, therefore, are resistant to the entanglement caused by the environment). 
The less the states entangle, the more stable they are. All other states 
evolve into joint system-environment states, preserving their purity.

Our concern in this Letter is to analyze the effect of the zero
point fluctuations of the environment, as a source of decoherence.
The coupling of a quantum system to an environment generally leads
to energy fluctuations in the test particle even at zero temperature
\cite{nagaevEPL}. Since phases are time integrals of energy, zero
point energy fluctuations make possible that decoherence occurs even
at zero temperature. These fluctuations are a consequence of the
finite coupling energy between the test particle (system) and the
bath, and of the fact that the Halmiltonian of the isolated system
does not commute with the interaction Hamiltonian.

Vacuum fluctuations have several observable effects. The Lamb shift
is a widely known example. Another one is the Casimir effect. In
these examples the effect of vacuum can be thought in terms of the
renormalization of the original parameters characterizing the
system. In contrast, the fluctuations we deal with in this Letter
are not only absorbed into renormalized parameters of the test
particle. Not only does the environment renormalize, but also it is
a source of dissipation and noise for the system. Therefore, we are 
considering the effect of quantum fluctuations of the environment 
over a quantum system, as the only source of decoherence (meaning that
there is no possible thermal fluctuations inducing classicality).

The question about the influence of zero temperature environment on
the interference phenomena has been discussed in the last years
\cite{ford,imry,sinha}. There have been studies on the temperature
dependent weak localization measurements \cite{4ratchov}, reporting
residual decoherence in metals at zero temperature, in contradiction
to theoretical predictions \cite{5ratchov}, and on the zero-point
decoherence induced by Coulomb interactions in disordered electron
systems; just to mention a few examples.

In previous works (see for example \cite{leshouches} for an
excellent review of the {\it state of the art} of decoherence) about
decoherence in quantum Brownian motion, most of the conclusions are
simply numerical or analytical only in the high temperature limit
\cite{jpphabzurek}. Low temperature case was discussed in
\cite{leshouches,jppdavila} showing a numerical estimation of the decoherence
rate. S. Sinha, in Ref. \cite{sinha}, studied the zero temperature
case analytically. Under some approximations the author found an
expression for the time dependence of the off-diagonal terms of the
density matrix. In this Letter, we complete that study showing an
exact calculation of diffusive terms, and also providing the
decoherence timescale for different situations of interest. In
addition, we solve the master equation for an upside-down Brownian
particle to emphasize the role of zero temperature fluctuations
during a second order phase transition \cite{guthpi,order}.


\section{The master equation at T=0}

Let us consider a quantum particle (characterized by its mass $M$
and its bare frequency $\Omega$) linearly coupled to an environment
composed of an infinite set of harmonic oscillators (of mass $m_n$
and frequency $\omega_n$). We may write the total action
corresponding to the system-environment model as (we set $\hbar =
1$)
\begin{eqnarray}S[x,q_n] &=& S[x] + S[q_n] + S_{\rm int}[x,q_n]\nonumber \\
&=& \int_0^t ds \left[\frac{1}{2} M ({\dot x}^2 - \Omega^2 x^2) +
\sum_n \frac{1}{2} m_n ({\dot q}_n^2 - \omega_n^2 q_n^2)\right] -
\sum_n C_n x q_n,\end{eqnarray} where $x$ and $q_n$ are the
coordinates of the particle and the oscillators, respectively. The
particle is coupled linearly to each oscillator with strength
$C_n$.

The relevant objects to analyze the quantum to classical transition
in this model are the reduced density matrix (obtained from the 
full density matrix integrating out all the degrees of 
freedom of the environment noted as ${\bar q}$), and the associated
Wigner function
\begin{eqnarray}
\rho_{\rm r} (x,x',t)&=& \int d{\bar q} \,\, 
\rho (x,{\bar q},x',{\bar q},t)\nonumber\\
W_{\rm r} (x,p,t)&=& \frac{1}{2\pi} \int_{-\infty}^{+\infty} dy~
e^{ipy} ~ \rho_{\rm r}(x+ \frac{y}{2}, x-\frac{y}{2},t).
\end{eqnarray}
The reduced density matrix satisfies a
master equation.
Hu-Paz-Zhang \cite{hpz} have evaluated the master equation for the quantum
Brownian motion problem (alternatively, one can write an equation of the Fokker-Planck type
for the reduced Wigner function \cite{jpphabzurek} in order to study the dynamics in
phase space)

\begin{equation}   \dot{\rho}_{\rm r} =-i\bigl[{H}_{\rm
syst}+ \frac{1}{2}M {\tilde \Omega}^2 x^2 ,\rho_{\rm r}\bigr]
- i\gamma(t)\bigl[x,\bigl\{ p,\rho_{\rm r}\bigr\}\bigr]
-D(t)\bigl[x,\bigl[ x,\rho_{\rm r}\bigr]\bigr]
 -f(t)\bigl[x,\bigl[ p,\rho_{\rm r}\bigr]\bigr].
\label{master}\end{equation}
The time dependent coefficients (in the case of weak coupling to the 
bath) are given by
\begin{eqnarray}
{\delta\Omega}^2(t) &=& - \frac{2}{M}\int_0^t ds \cos(\Omega s)
\eta(s)\nonumber \\
\gamma(t) &=& \frac{1}{M\Omega}\int_0^t ds \sin(\Omega s)
\eta(s)\nonumber
\\
D(t) &=& \int_0^t ds \cos(\Omega s) \nu(s)\label{coef} \\
f(t) &=& -\frac{1}{M\Omega}\int_0^t ds \sin(\Omega s)
\nu(s),\nonumber
\end{eqnarray}
where ${\delta\Omega}^2(t)$ is the shift in frequency, which produces the 
renormalized frequency ${\tilde\Omega}^2$ that appears in the master equation. 
$\gamma (t)$ is the dissipation coefficient related to the friction kernel 
defined below, and $D(t)$
and $f(t)$ are the diffusion coefficients, which produce the
decoherence effects. Diffusion coefficients come from the noise kernel, source 
of stochastic forces in the associated Langevin equation. $f(t)$ is named 
anomalous in the literature since it generates a second derivative term in 
the phase space representation of the evolution equation, just like the 
ordinary diffusion term \cite{unruhzu}. $\eta (t)$ and $\nu (t)$ are the dissipation
and noise kernels, respectively,
\begin{eqnarray}\eta (t)& =& \int_0^\infty d\omega I(\omega ) \sin \omega t
\nonumber \\
\nu (t) &=& \int_0^\infty d\omega I(\omega ) \coth \frac{\beta
\omega}{2} \cos \omega t\nonumber, \nonumber\end{eqnarray}
$I(\omega )$ is the spectral density of the environment,
defined as $I(\omega ) = (2/\pi) M\gamma_0 \Lambda^2
\omega/(\omega^2 + \Lambda^2)$ (where $\Lambda$ is the physical high-frequency cutoff, 
which represents the highest frequency present in the environmet). In the
high temperature limit of an ohmic environment (where $I(\omega ) \propto \omega $) the
coefficients in Eq.(\ref{coef}) become constants. In particular,
the diffusion coefficient can be approximated by $D \simeq 2
\gamma_0 k_{\rm B} T M$, where $\gamma_0$ is the dissipation
coefficient \cite{hpz}. In this limit, while $\gamma_0$ is a
constant and $D \propto T$, the coefficient $f
\propto T^{-1}$ can be neglected. Therefore, the term proportional to $D$ is the
relevant one in the master equation at high temperatures in order to
evaluate decoherence.

We will evaluate the time dependent coefficients of the master
equation at zero temperature. For this, we set $\coth
\beta\hbar\omega/2 = 1$. These coefficients have been evaluated 
previously in Refs.\cite{jppdavila,vinales}.

The shift in frequency ${\delta\Omega}^2 (t)$ is,

\begin{equation}
{\delta\Omega}^2 (t) = -\frac{4\gamma_0}{\pi}\Lambda^2 \int_0^\infty
d\omega \int_0^t ds \frac{\omega}{\omega^2 + \Lambda^2} \sin\omega
s \cos\Omega s,
\label{Omega}\end{equation}
performing integrations, we obtain

\begin{equation}
{\delta\Omega}^2 (t) = -2 \gamma_0 \frac{\Lambda^3}{\Lambda^2 +
\Omega^2} \left[1 - e^{-\Lambda t} \left(\cos\Omega t -
\frac{\Omega}{\Lambda} \sin\Omega t\right)\right],
\label{Omega2}\end{equation} for times such that $\Lambda t > 1$ the
shift reads (see Fig. 1 (a))

\begin{equation}
{\delta\Omega}^2  = -2 \gamma_0 \frac{\Lambda^3}{\Lambda^2 +
\Omega^2}.\label{Omegalarge}
\end{equation}

Dissipation coefficient (Fig. 1 (b)) comes from the integral

\begin{equation}
\gamma (t) = \frac{2\gamma_0}{\pi\Omega}\Lambda^2 \int_0^\infty
d\omega \int_0^t ds \frac{\omega}{\omega^2 + \Lambda^2} \sin\omega
s \sin\Omega s,\label{gamma}
\end{equation}
and it is given by

\begin{equation}
\gamma (t) =  \gamma_0 \frac{\Lambda^2}{\Lambda^2 + \Omega^2}
\left[1 - e^{-\Lambda t} \left(\cos\Omega t + \frac{\Lambda}{\Omega}
\sin\Omega t\right)\right], \label{gamma2}\end{equation} which has
the following asymptotic behavior

\begin{equation}
\gamma (t) =  \gamma_0 \frac{\Lambda^2}{\Lambda^2 + \Omega^2}.
\label{gammalarge}\end{equation}

The normal diffusive coefficient (normally connected with
decoherence effects) is coming from the integral

\begin{equation}
D(t)= \frac{2M\gamma_0}{\pi}\Lambda^2 \int_0^\infty d\omega \int_0^t ds
\frac{\omega}{\omega^2 + \Lambda^2} \cos\omega s \cos\Omega s.
\label{D}\end{equation}

This integral can be exactly solved. The result is:

\begin{eqnarray}
D(t)&=&\frac{2M\gamma_0}{\pi} \frac{\Lambda^2 \Omega}{\Omega^2 + \Lambda^2}
\left[Shi(\Lambda t) \left(\frac{\Lambda}{\Omega}\cos\Omega t
\cosh\Lambda t +
\sin\Omega t \sinh\Lambda t\right)\right. \nonumber \\
&-& \left.
Chi(\Lambda t) \left(\frac{\Lambda}{\Omega}\cos\Omega t \sinh\Lambda t +
\sin\Omega t \cosh\Lambda t\right) \right. \nonumber \\
&+& \left. Si(\Omega t) \right],
\label{D2}\end{eqnarray}
where, $Chi(x)$ and $Shi(x)$ are the
hyperbolic CosIntegral and SinIntegral respectively;
$Si(x)$ is the SinIntegral.

The expression can be very well approximated, when $\Lambda t > 1$,
by

\begin{equation}
D(t) = \frac{2M\gamma_0}{\pi} \frac{\Lambda^2 \Omega}{\Omega^2 + \Lambda^2}
Si(\Omega t).
\label{Dlarge}\end{equation}
This coefficient is the normal diffusion at $T=0$. This is for any
value of $\Omega$. We are interested in time scales longer than
the memory time $1/\Lambda$. This coefficient is an oscillatory
function of time. In fact, only in the limit  $\Omega t \gg 1$, as
$Si$ goes to $\pi/2$, we can have an asymptotic value $D_\infty
\sim M \gamma_0 \Lambda^2 \Omega/(\Lambda^2 + \Omega^2)$,
independent of time \cite{leshouches} (Fig. 1 (e)). In any other case, the 
coefficient has an
initial transient and approaches the asymptotic value $D_\infty$
as the $Si$ (see Fig. 1 (c)). It is important to note, that in the
opposite case, when $\Omega t \ll 1$, normal diffusion is a
linearly growing function of time, $D \sim 2M(\gamma_0/\pi)
\Lambda^2\Omega^2 t/(\Lambda^2 + \Omega^2)$, similar to the result
obtained in Ref.\cite{sinha}.

The anomalous diffusion coefficient is given by the integral

\begin{equation}
f(t)= \frac{
2\gamma_0}{\pi\Omega}\Lambda^2 \int_0^\infty d\omega \int_0^t ds
\frac{\omega}{\omega^2 + \Lambda^2} \cos\omega s \sin\Omega s.
\label{f}\end{equation}
It reads as

\begin{eqnarray}
f(t)&=& 2\gamma_0 \frac{\Lambda^2}{\Omega^2 + \Lambda^2}
\left[Shi(\Lambda t) \left(\frac{\Lambda}{\Omega}\sin\Omega t
\cosh\Lambda t -
\cos\Omega t \sinh\Lambda t\right)\right. \nonumber \\
&+& \left.
Chi(\Lambda t) \left(- \frac{\Lambda}{\Omega}\sin\Omega t \sinh\Lambda t +
\cos\Omega t \cosh\Lambda t\right) \right. \nonumber \\
&-& \left. Ci(\Omega t) - \log\frac{\Lambda}{\Omega} \right].
\label{f2}\end{eqnarray}
Again, for $\Lambda t > 1$, this coefficient can be written as (Fig. 1 (d))

\begin{equation}
f(t) = 2\gamma_0 \frac{\Lambda^2}{\Omega^2 + \Lambda^2}
\left(-Ci(\Omega t) - \log\frac{\Lambda}{\Omega}\right),
\label{flarge}\end{equation}
coefficient $f(t)$ also approaches an asymptotic
value when $\Omega t \gg 1$, $f_\infty \sim -2\gamma_0 (\Lambda^2/(\Lambda^2 + \Omega^2)) 
\log\Lambda/\Omega$ (Fig. 1 (f)); and it does to 
$f(t) \sim - 2 \gamma_0 (\log\Lambda t + \Gamma )$, 
when $\Omega t \ll 1$ ($\Gamma$ is the EulerGamma number).

\begin{figure}[t]
\includegraphics[width=10cm]{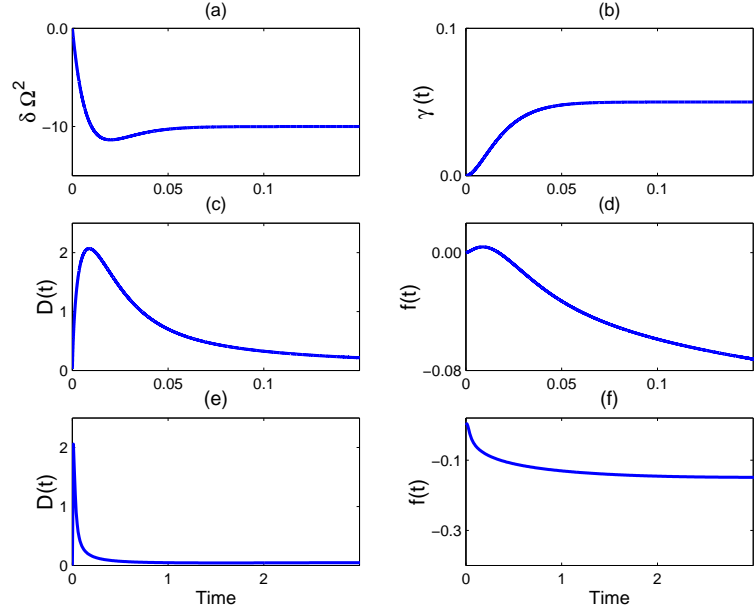}
\caption{Time dependence of the coefficients of the master equation
at $T=0$. On top we show frequency renormalization (a) and
dissipation coefficients (b). Plots below show the normal $D$ (c)
and anomalous $f$ (d) diffusion coefficients for
short times in order to show the initial transient. Asymptotic
values of diffusion are shown in (e) and (f). The parameters
used in the plot are $\gamma_0 = 0.05$, $\Lambda = 100$, $\Omega =
1$.} \label{figure1}
\end{figure}

In Fig. 1, we show the time behavior of these coefficients. It is easy to probe
asymptotic behavior also from analytical expressions.

It has been noted that the master equation in the high temperature limit 
(or even in the $\gamma \sim$ constant approximation) has the pathology that 
the density matrix loses its positivity at short 
times (shorter than $1/k_BT$). This violation is essentially due to the action 
of the friction term. Master equation (\ref{master}) does not have 
the pathological behaviour of the master equation at high temperature \cite{hpz}. 
The dissipation coefficient is a time-dependent function that vanishes initially 
together with its first derivative. Therefore, the initial behaviour of 
the density matrix is diffusion dominated and positivity is preserved, even 
in the perturbative case, up to second order with respect to the 
coupling constant between system and environment.

With these coefficients at hand, we will evaluate the decoherence
time following Refs.\cite{leshouches,jpphabzurek}.

\section{Decoherence time at T = 0}

We will analyze the decoherence process in a simple case. We prepare
an initial superposition of delocalized (in position or momentum)
states. We consider two wave packets symmetrically located in phase
space, of the form \cite{jpphabzurek}: 
$\Psi(x,t=0) = \Psi_1(x) + \Psi_2(x)$,
where
\begin{equation}
\Psi_{1,2} = N \exp\left(-\frac{(x\mp L_0)^2}{2\delta^2}\right)
\exp(\pm i P_0x),\end{equation}
\begin{equation} N^2 = \frac{{\tilde N}^2}{\pi\delta^2}=
\frac{1}{2\pi\delta^2}\left[1 + \exp\left(-\frac{L_0^2}{\delta^2}
- \delta^2 P_0^2\right)\right]^{-1},\end{equation}
where $N$ is normalization, and $\delta$ is the initial width of the 
wave packet.
In terms of the Wigner function, the state at time $t$ is
$W(x,p,t) = W_1(x,p,t) + W_2(x,p,t) + W_{\rm int}(x,p,t)$, 
where
\begin{equation}W_{1,2}= \frac{{\tilde
N}^2}{\pi}\frac{\delta_1}{\delta_2}\exp\left(-\frac{(x\mp
x_c)^2}{\delta_1^2}\right)\exp\left(-\delta_2^2(p\mp p_c - \beta
(x \mp x_c))^2\right),
\end{equation}
and
\begin{equation}W_{\rm int}=\frac{2{\tilde
N}^2}{\pi}\frac{\delta_1}{\delta_2}\delta_2^2 (p - \beta x)^2
\cos(2k_p p + 2 (k_x - \beta k_p)x).\end{equation}

All the coefficients are functions of time, determined
by the evolution propagator of the reduced density matrix and the
initial state. The explicit form can be found in
Ref. \cite{jpphabzurek}. The initial state is such that
$\delta_1^2=\delta_2^2=\delta^2$, $k_x = P_0=p_c$, $k_p=L_0=x_c$.
$k_p$ and $k_x$ indicate the evolution of the fringes in the 
momentum and coordinates directions of the phase space.

As it was defined in the previous literature (see for example
\cite{leshouches}), the effect of decoherence is produced by an
exponential factor $\exp(-A_{\rm int})$, defined as

\begin{equation}
\exp(-A_{\rm int}) = \frac{1}{2}\frac{W_{\rm int}(x,p)|_{\rm
peak}}{\left[W_1(x,p)|_{\rm peak} W_2(x,p)|_{\rm
peak}\right]^{\frac{1}{2}}}.\end{equation}

Initially, $A_{\rm int}=0$, and it is always bounded $A_{\rm int}
\leq L_0^2/\delta^2 + \delta^2 P_0^2 = A_{\rm int}|_{\rm max}$.
The fringe visibility factor $A_{\rm int}$ evolves in time
as
$
{\dot A}_{\rm int}= 4 D(t) k_p^2 - 4 f(t) k_p (k_x - \beta
k_p)$. In the high temperature approximation, the
anomalous coefficient is neglected and we obtain the very well known
decoherence rate considering only the constant diffusion term,
proportional to $T$. In our present case, at zero temperature, both
coefficients $D$ and $f$ contribute to the fringe visibility factor.
A conservative choice is to assume fringes always stay more or less 
frozen at the initial values, and we can set $k_p=L_0$ and $k_x=1/(2L_0)$. 
Neglecting the initial transient (i.e. $\Lambda t > 1$), we use Eqs.(\ref{Dlarge}) and
(\ref{flarge}) to evaluate $A_{\rm int}$. In order to have the
simplest analytical expression for the decoherence rate, we use 
a short time approximation to evaluate $\beta$, giving $\beta \sim 0$ 
(see \cite{jpphabzurek}). Thus, we get

\begin{equation}
\dot{A}_{\rm int}\approx 4 L_0^2 D(t) - 2 f(t) 
.\label{aint}\end{equation}

In order to evaluate the decoherence time $t_D$, we have to solve $1
\approx A_{\rm int}(t = t_D)$. From Eq.(\ref{aint}) it is not possible 
to find a global decoherence time-scale at $T = 0$. Nevertheless,
we can find limits in which we are able to give different scales for
decoherence.

For example, for large natural frequency $\Omega$, such
as $\Omega \sim \Lambda$ ($\Omega t \gg 1$), it is easy to see that 

\begin{equation} A_{\rm int} \sim 2 L_0^2 M \gamma_0 \Lambda t +
 4 \gamma_0 \left(t ~ Ci(\Lambda t) - 
\frac{\sin \Lambda t}{\Lambda}\right),
\end{equation}
giving a very short decoherence time-scale,

\begin{equation}
t_D \sim \frac{1}{2 M L_0^2 \gamma_0 \Lambda}.
\label{td1}\end{equation}
This result will be valid as long as the product $M L_0^2 \gamma_0
\leq 1$, in order to be able to neglect the initial transient. It
coincides with the decoherence time evaluated directly from
$D_\infty$ as in Ref.\cite{leshouches}. In this limit, the anomalous
coefficient does not play any role (as we could check using $f_\infty$ 
in (\ref{td1})).

In the opposite limit, when $\Omega t \ll 1$ 
(for times $\frac{1}{\Lambda} < t < \frac{1}{\Omega}$), we can
approximate $A_{\rm int}$ using asymptotic limit of $Si$ and 
$Ci$ by

\begin{equation}
A_{\rm int} \approx \frac{8\Lambda^2}{\Lambda^2 + \Omega^2} \gamma_0 \left[
\frac{M L_0^2}{2\pi} (\Omega t)^2 + t ~ (\log\Lambda t + \Gamma - 1)\right]  ,
\label{a2}\end{equation}
resulting in a decoherence time bound, $t_D \leq \frac{1}{8  \gamma_0}$,
which could be large for very underdamped systems. Here, the
logarithmic correction is due to the $f(t)$ diffusion term (unlike Ref. \cite{sinha}, 
where anomalous diffusion was neglected). This  
scale is longer than the decoherence time in the high temperature, even 
in the case of low temperature for high natural frequency of the Brownian 
particle. Our result is still smaller than the saturation time $t_{\rm sat} 
= \gamma_0^{-1}$, the time in which $A_{\rm int}$ reaches its maximum 
value.

In the case we can neglect the second term in (\ref{a2}) [for example 
considering ``macroscopic'' trajectories (large $M L_0$)], we can show

\begin{equation}
t_D \approx \frac{1}{2L_0\Omega}\sqrt{\frac{\pi}{M\gamma_0}}.
\end{equation}

Summarizing, in this Section we have shown analytical expressions 
for the decoherence rate at zero temperature. We were able to
extract the decoherence timescales in different cases,
giving new results respect to previous works
\cite{leshouches,sinha} and showing how to get known numerical results.


\section{Decoherence for the upside-down harmonic oscillator}

In this Section we are concerned with the analysis of the quantum to
classical transition of the order parameter during a second order
phase transition \cite{order}. In a realistic model one should
address this problem in the context of quantum field 
theory\cite{ray}. This is a very difficult task since non Gaussian and
non perturbative effects are relevant. For this reason, we will only
concentrate here in a toy model in ordinary quantum mechanics.

Guth and Pi \cite{guthpi} considered an upside down harmonic
oscillator as a toy model to describe the quantum behavior of this
unstable system. This toy model should be a good approximation for
the early time evolution of the phase transition, as long as one can
neglect the non-linearities of the potential \cite{order}. In this
Section, we will analyze the decoherence effects during a quantum
phase transition in which the environment is at $T=0$.

Let us consider the unstable quantum particle (characterized by its
mass $M$ and its bare frequency $\Omega$) linearly coupled to a zero
temperature environment composed of an infinite set of harmonic
oscillators (of mass $m_n$ and frequency $\omega_n$). As the
coupling between system and environment is lineal, the result is
exact, and can be easily obtained by replacing $\Omega$ by $i
\Omega$ in the Hu-Paz-Zhang equation. If the initial  wave function
is Gaussian, it will remain Gaussian for all times (with time
dependent parameters that set its amplitude and spread).

Let us solve Eqs.(\ref{master}) using a
Gaussian ansatz for the reduced density matrix

\begin{equation}\rho_r (\Sigma ,\Delta ,t) = N(t) e^{-(2 a - C) \Delta^2}
e^{- (2 a + C) \Sigma^2} e^{- 4 i b \Sigma \Delta},\label{newrho}
\end{equation}
while the reduced Wigner function is exactly evaluated as
\begin{equation}W_r(x,p,t) = \frac{1}{\pi} \sqrt{\frac{2 a + C}{2 a - C}}
e^{-( 2 a + C) x^2} e^{- \frac{(p - 2 x b)^2}{(2 a -
C)}},\label{neww}\end{equation}
where $C(t)$ is a real function; and where $\Sigma = 1/2(x+x')$ and $\Delta = 1/2(x -
x')$.

The master equation, in the zero
temperature limit, becomes
\begin{eqnarray}\dot{a} &=& 4 a b - \gamma (t) (2 a - C)+ D(t) + 2
b f(t)
\nonumber \\
\dot{b} &=& - 2 (a^2 - 2 b^2 - \frac{C^2}{2}) -
\frac{1}{2} (\Omega^2 + {\delta\Omega}^2)  - 2 b \gamma (t) - (2 a - C) f(t) \nonumber \\
\dot{C} &=& 4 C b + 2 (2 a - C)\gamma (t)  - 2 D(t) - 4 b f(t)
 \nonumber \\
\dot{N} &=& 2 N b. \label{coopereqs}\end{eqnarray} In this case,
the temporal coefficients are given by (in the $\Lambda t > 1$
limit)

\begin{eqnarray}
{\delta\Omega}^2 = -\frac{2\gamma_0\Lambda^3}{\Lambda^2 - \Omega^2} &;& \gamma
=\frac{\gamma_0\Lambda^2}{\Lambda^2 - \Omega^2}\nonumber \\
D(t)= \frac{2M\gamma_0}{\pi}\frac{\Omega\Lambda^2}{\Lambda^2 -
\Omega^2}Shi(\Omega t) &;& f(t) =
-\frac{2\gamma_0\Lambda^2}{\Lambda^2 - \Omega^2}\left(Chi(\Omega
t) + \log\frac{\Lambda}{\Omega}\right).\nonumber \end{eqnarray}

From Eqs.(\ref{newrho}) and (\ref{neww}) we see that the relevant
function to describe correlations and decoherence is now $2a-C$. For
$2a - C =O(1)$ we have both correlations and decoherence. The set of
Eqs.(\ref{coopereqs}) can be solved numerically. In Fig. 3 we show
the behavior of $2a-C$ as a function of time. We see that it tends
asymptotically to a constant of order one (of course the asymptotic
value depends on the properties of the environment).

\begin{figure}[t]
\includegraphics[width=6cm]{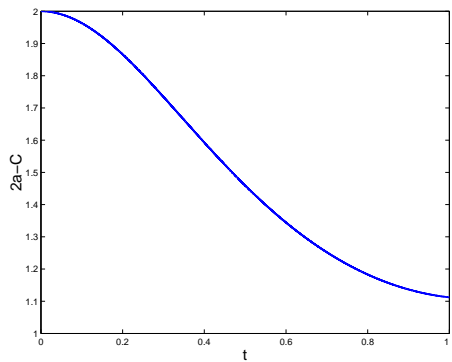}
\caption{This coefficient measures the importance of the
nondiagonal terms in the density matrix. Alternatively, it is the
 width of the Wigner function. Is shows the rapid decoherence
 of the unstable particle coupled to a zero temperature environment.
 Parameters are $\gamma_0 = 0.01$,
 $\Omega = 1$, and $\Lambda = 100$.} \label{figure3}
\end{figure}

The main conclusion of this Section is the following. In order to
study a sudden quench quantum phase transition, at early times we
can use the upside down potential \cite{ray,guthpi,raynpb}. When the system
is isolated, due to the high squeezing of the initial wave packet,
$x$ and $p$ become classically correlated \cite{order}. The density
matrix is not diagonal. The ``correlation time" depends on the shape
of the potential. However, when the particle is coupled to an
environment, a true quantum to classical transition takes place. The
Wigner function becomes peaked around a classical trajectory and the
density matrix diagonalizes. The decoherence time at $T= 0$ depends
on the diffusion coefficients $D$ and $f$ and plays an important
role in the early stages of a quantum phase transition, inducing
classicality of the order parameter. Quantum aspect could be
relevant if non-linearities are taken into account \cite{nuno}.
Decoherence allows a classical description even in the nonlinear
regime.

\section{Acknowledgments}
This work was supported by UBA, CONICET, Fundaci\'on Antorchas and
ANPCyT, Argentina. We would like to thank D. Mazzitelli, D.
Monteoliva and J.P. Paz for useful conversations.


\begin{thebibliography}{99}

\bibitem{giul} See for example, {\it Decoherence and the appearance of
a classical world in quantum theory}, D. Giulini et al, Springer
Verlag (1996).

\bibitem{unruhzu}W.G. Unruh and W.H. Zurek, Phys. Rev. {\bf D40}, 1071
(1989).

\bibitem{zurek} W.H. Zurek, ``Prefered Sets of States, Predictability, Classicality, and
Environment-Induced Decoherence''; in {\it The Physical Origin of
Time Asymmetry}, ed. by J.J. Halliwell, J. Perez Mercader, and
W.H. Zurek (Cambridge University Press, Cambridge, UK, 1994).

\bibitem{nagaevEPL} K.E. Nagaev and M. Buttiker, Europhys. Lett.,
{\bf 58}(4), 475 (2002).

\bibitem{ford} G.W. Ford and R.F. O'Connell, J. Optics {\bf B5}, S349 (2003).

\bibitem{imry} Y. Imry, arXiv: cond-mat/0202044.

\bibitem{sinha} S. Sinha, Phys. Lett. {\bf A228}, 1 (1997).

\bibitem{4ratchov} P. Mohanty, E.M.Q. Jariwala, and R.A. Webb,
Phys. Rev. Lett. {\bf 78}, 3366 (1997).

\bibitem{5ratchov} B.L. Altshuler, A.G. Aronov, and D.E.
Khmelnitsky, J. Phys. C: Solid State Phys., {\bf 15}, 7367 (1982).

\bibitem{leshouches} J.P. Paz and W.H. Zurek, {\it
Environmet-induced decoherence and the transition from quantum to
classical}, lectures at the 72nd Les Houches Summer School on
"Coherent Matter Waves"; (1999). arXiv: quant-ph/0010011.

\bibitem{jpphabzurek} J.P. Paz, S. Habib, and W.H. Zurek, Phys. Rev.
 {\bf D47}, 488 (1993).


\bibitem{jppdavila} J.P. Paz and L. D\'avila Romero, Phys. Rev.
 {\bf A55}, 4070 (1997).

\bibitem{guthpi} A. Guth and S.Y. Pi, Phys. Rev {\bf D}32, 1899
(1985).

\bibitem{order} F.C. Lombardo, F.D. Mazzitelli, and D. Monteoliva,
Phys. Rev. {\bf D62}, 045016 (2000).

\bibitem{hpz} B.L. Hu, J.P. Paz, and Y. Zhang, Phys. Rev. {\bf D45}, 2843
(1993).

\bibitem{vinales} A.D. Vi\~nales, Tesis de Licenciatura en F\'\i sica, 
(M.Sc. in Physics); University of Buenos Aires, unpublished (2002).  

\bibitem{ray} F.C. Lombardo, F.D. Mazzitelli, and R.J. Rivers,
Phys. Lett. {\bf B523}, 317 (2001).

\bibitem{raynpb} F.C. Lombardo, F.D. Mazzitelli, and R.J. Rivers, 
Nucl. Phys. {\bf B672}, 462 (2003).

\bibitem{nuno} N.D. Antunes, F.C. Lombardo, and D. Monteoliva, Phys.
Rev. {\bf E64}, 066118 (2001).

\end{thebibliography}
\end{document}